\pdfoutput=1
\documentclass[conference]{IEEEtran}
\usepackage{cite}
\usepackage{tabularx,booktabs}
\newcolumntype{C}{>{\centering\arraybackslash}X} 
\ifCLASSINFOpdf
\usepackage[pdftex]{graphicx}
\DeclareGraphicsExtensions{.pdf}
\else
\fi
\usepackage[cmex10]{amsmath}
\usepackage{amsfonts}
\newcommand{\overbar}[1]{\mkern 1.5mu\overline{\mkern-1.5mu#1\mkern-1.5mu}\mkern 1.5mu}
\begin{document}
%
\title{Bit Partitioning Schemes for Multicell Zero-Forcing Coordinated Beamforming}

\author{\IEEEauthorblockN{Pranav A}
\IEEEauthorblockA{VIT University, Vellore, India\\ Email : cs.pranav.a@gmail.com}}


%


\maketitle

\begin{abstract}
In this paper, we have studied the bit partitioning schemes for the multicell multiple-input and single-output (MISO) infrastructure. Zero forcing beamforming is used to null out the interference signals and the random vector quantization, quantizes the channel vectors. For minimal feedback period (MFP), the upper bound of rate loss is calculated and optimal bit partitioning among the channels is shown. For adaptive feedback period scheme (AFP), joint optimization schemes of feedback period and bit partitioning are proposed. Finally, we compare the sum rate efficiency of each scheme and conclude that minimal feedback period outperforms other schemes.
\end{abstract}


%
\IEEEpeerreviewmaketitle

\section{Introduction}

Recent works of communication system designs in channel model designing make use of adaptive routines which determine the desired parameters in order to increase the spectral efficiency of the model \cite{mimo}\cite{beamforming}. The advantage of these approaches over the static techniques is that they adjust accordingly to the channel propagation conditions. Here we make use of multiple antenna broadcast channels to study these techniques. These broadcasting channels have an edge over the single antenna channels like improved rate, increased diversity, beamforming and interference cancelling abilities \cite{overview}. 

The transmitter makes use of channel state information (CSI) to model the appropriate channel conditions. The feedback is given by the receiver in the form of low-rate data uplink, which is also known as \textit{limited feedback}. Thus by making the use of the limited feedback, the transmitter updates CSI accordingly \cite{update}.

Coordinated beamforming \cite{comp} is commonly used to allocate suitable bits to a particular channel. There are many ways to quantize the CSI vectors \cite{Jindal}, however we use Random Vector Quantization (RVQ) here \cite{rvq} for a more receptive facile analysis. This study focuses on two adaptive bit allocation schemes:
\begin{enumerate}
\item \textbf{Minimal Feedback Period (MFP)} : Here, only bits for each channel are optimized. The feedback is given every time. \cite{MFP}
\item \textbf{Adaptive Feedback Period (AFP)} : Here, bits for each channel and feedback update period are optimized. The feedback is given after a certain amount of time. \cite{awful}
\end{enumerate}

The objective of this study is to: 
\begin{itemize}
\item analyze and derive closed-form expression for MFP using Gauss-Markov model.
\item formulate joint optimization of parameters for AFP.
\item simulate the bit allocation schemes.
\end{itemize}

\textbf{Notations} : $\mathbf{M}$ denotes the matrix and $\mathbf{m}$ denotes the vector. $\mathbf{M}^{\mathit{H}}$ stands for the hermitian transpose. $\mathbb{E} (\cdot)$ stands for the expectation operator. $|\mathbf{m}|$ stands for the 2-norm of a complex vector. $\mathcal{N}(\mu, \sigma^2)$ denotes the Gaussian-distribution with mean $\mu$ and variance $\sigma^2$.

The paper is organized into the following sections : Section 2 gives the details of the system model. Section 3 lays down the mathematical definitions of interference calculations with perfect and delayed CSI. The derivation of the closed-form expression of MFP is explicated in section 4. Section 5 demonstrates the joint optimization process of AFP. Finally simulation results are shown in section 6.

\section{System Model}

\begin{figure}[!t]
\centering
\includegraphics[width=3.3 in]{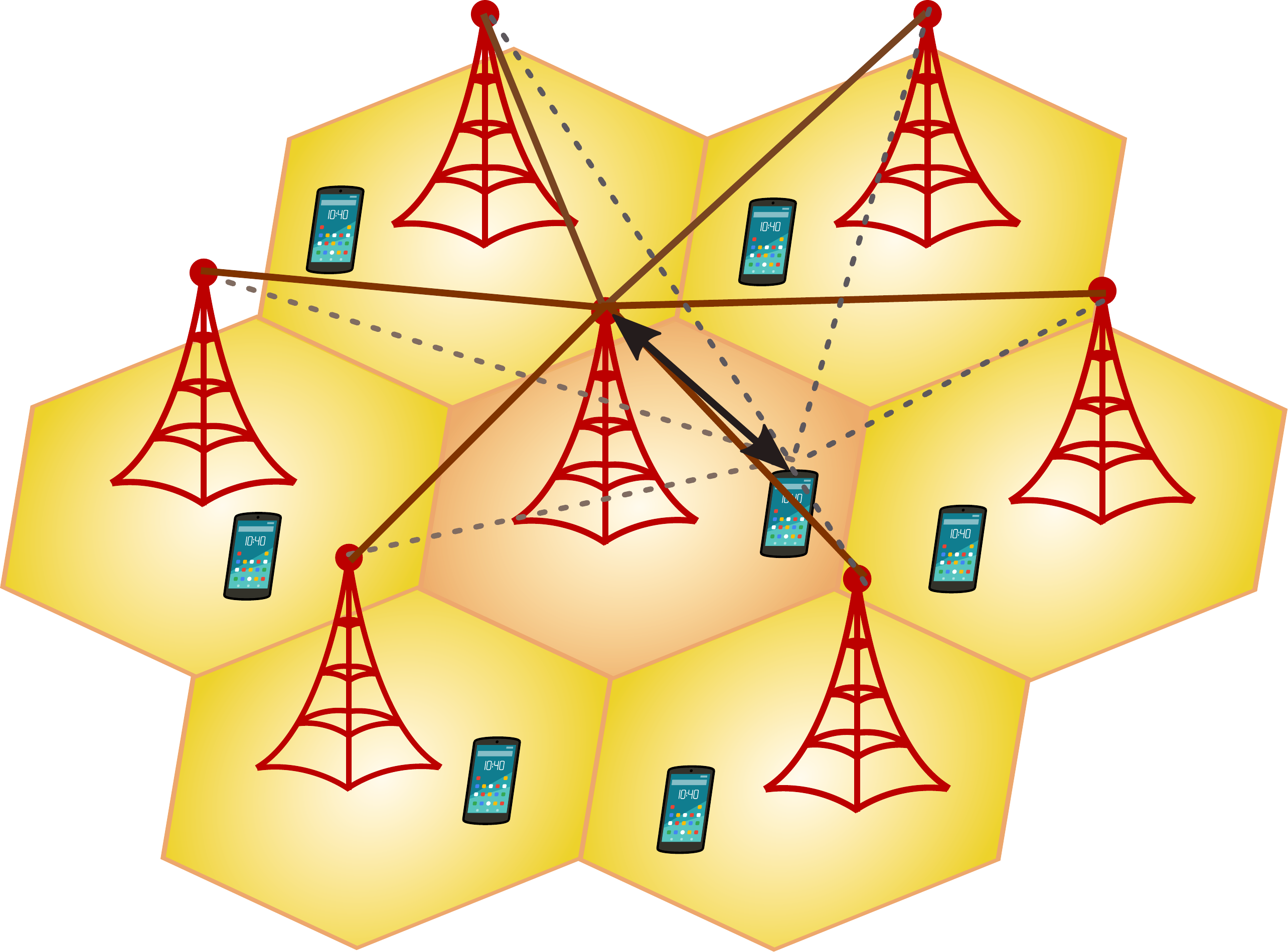}
\caption{Simplified sketch of the system model. Towers depict the base stations in a cell and mobiles represent users' mobile stations. Arrowed line between the base station and user is the desired signal. The dashed lines are interference signals and solid lines between the stations specify the channel station information.}
\label{model}
\end{figure}

The simplified system model is depicted in Figure \ref{model}. Here we model the MISO (multiple-input single-output) scenario in a time varying channel. We assume that a base station (BS) has $M$ antennas and it is exclusive to a particular user for a certain period of time. We assume that the network consists of $K$ cells and the base station will be interfered by $K - 1$  channels. $\mathbf{h}_{\mathit{ij}} [\mathit{n}]$ denotes the channel vector of the $i^{th}$ base station serving the user at the $j^{th}$ station at the $n^{th}$ instant. Here, $\mathbf{h}_{\mathit{ij}} [\mathit{n}] \in \mathbb{C}^M$, where $\mathbb{C}$ is a set of complex numbers. Thus $\mathbf{h}_{\mathit{ij}} [\mathit{n}] \, \forall \, i = j$ is the desired channel vector and $\mathbf{h}_{\mathit{ij}} [\mathit{n}] \, \forall \, i \neq j$ is the interference vector.

We denote the beamforming \cite{sinr} vector of unit norm for the station $i$ as $\mathbf{b}_{\mathit{i}} [\mathit{n}]$. We assume that $\mu_{ij}$ is the signal power constraint for station $i$ to user at station $j$.

Thus, (SINR) the signal to interference ratio \cite{Jindal} of the $\mathit{i}^{th}$ user at the $\mathit{n}^{th}$ instant is given by:

\begin{equation}
SINR_{\mathit{i}} [\mathit{n}] = \frac{ \mu_{ii} | \mathbf{h}_{\mathit{ii}}^{\mathit{H}} [\mathit{n}] \mathbf{b}_{\mathit{i}} [\mathit{n}] |^2 }{ 1 + \sum_{j = 1, j \neq i}^{K} \mu_{ij}  |\mathbf{h}_{\mathit{ij}}^{\mathit{H}} [\mathit{n}] \mathbf{b}_{\mathit{j}} [\mathit{n}] |^2}
\end{equation}

The sum rate of all the users is given as:

\begin{equation}
R_{s} [\mathit{n}] = \sum_{j = 1}^{K} \log_2 (1 + SINR_{\mathit{j}} [\mathit{n}])
\end{equation}

We have used first order Gauss-Markov process \cite{markov} to model the channel vectors in a time varying channel.

\begin{equation}
\mathbf{h}_{\mathit{ij}} [\mathit{n}] = \epsilon_{\mathit{ij}} \mathbf{h}_{\mathit{ij}} [\mathit{n} - 1] + \sqrt{1 - \epsilon_{\mathit{ij}}^2} \mathbf{w}_{\mathit{ij}} [\mathit{n}]
\end{equation}

where $\mathbf{w}_{\mathit{ij}} [\mathit{n}] \in \mathbb{C}^M \text{ and } \mathcal{N}(0, 1)$ and $\epsilon_{\mathit{ij}}$ is the correlation coefficient between station $i$ and station $j$, which is defined by the Clarke's \cite{clarke} model:

\begin{equation}
\epsilon_{\mathit{ij}} = \mathit{J}_0 (2 \mathit{\pi f_d^{ij} t})
\end{equation}

Here, $\mathit{J}_0$ is the Bessel function of zeroth-order, $t$ is the time-interval of a subframe and $f_d^{ij}$ is the maximum Doppler frequency, which is $f_d^{ij} = \frac{v_{ij} f_c}{c}$. Here $v_{ij}$ is the relative velocity between $i^{th}$ station to the $j^{th}$ user, $f_c$ is the carrier frequency and $c$ is the speed of light.

\section{Inter-cell Interference Calculations}

\subsection{Perfect CSI Modelling}

We assume that base station \textit{j} has correct interference channel state information (CSI). The norm of channel vector can be defined as:

\begin{equation}
\mathbf{\bar{h}}_{\mathit{ij}} [\mathit{n}] = \dfrac{\mathbf{h}_{\mathit{ij}} [\mathit{n}]}{\| \mathbf{h}_{\mathit{ij}} [\mathit{n}] \|} \text{ for } \mathit{i} \in \{1, 2, \dots, \mathit{K}\} \quad \forall \mathit{i} \neq \mathit{j}
\end{equation}

We use the zero-forcing method to calculate beamforming vector which nulls out the interference channels, which is defined as:

\begin{equation}
\mathbf{\bar{H}}_{\mathit{i}} [\mathit{n}] \mathbf{b}_{\mathit{i}} [\mathit{n}] = 0 \text{ for } \mathit{i} \in \{1, 2, \dots, \mathit{K}\}
\end{equation}

where $\mathbf{\bar{H}}_{\mathit{i}} [\mathit{n}] \in \mathbb{C}^{(K - 1) \times M}$ be representation of interference channel $\mathit{i}$. This is given by

\begin{equation}
\mathbf{\bar{H}}_{\mathit{i}} [\mathit{n}] =  \left[\mathbf{\bar{h}}_{1 \mathit{i}} [\mathit{n}], \dots, \mathbf{\bar{h}}_{\mathit{(i - 1)i}} [\mathit{n}], \mathbf{\bar{h}}_{\mathit{(i + 1)i}} [\mathit{n}], \dots, \mathbf{\bar{h}}_{\mathit{Ki}} [\mathit{n}] \right]^{\mathit{H}}
\end{equation}

Thus, the nullity of the matrix $\mathbf{\bar{H}}_{\mathit{i}} [\mathit{n}]$ is $M - K + 1$. For $M = K$, nullspace of the interference matrix will have a unique direction. The span of the nullspace is calculated through singular value decomposition. Let $\mathbf {M} =\mathbf {U} {\boldsymbol {\Sigma }}\mathbf {V} ^{*}$ be the singular value decomposition of the matrix $\mathbf {M}$. Let $\lambda_i$ be the eigenvalue in $\boldsymbol {\Sigma }$ and $\mathbf{x}_i$ be the corresponding eigenvector to $\lambda_i$. We know from the definition of the eigenvalues that, $\mathbf{M}\mathbf{x}_i = \lambda_i \mathbf{x}_i$. If we pick the $\lambda_i \approx 0$, then the equation reduces into $\mathbf{M}\mathbf{x}_i \approx 0$. Thus we can say that $\mathbf{x}_i$ is the span of the nullspace of $\mathbf{M}$. Through this method, the nullspace of the $\mathbf{\bar{H}}_{\mathit{i}} [\mathit{n}]$  is calculated. 

The resulting SINR will be:

\begin{equation}
SINR_{\mathit{i}} [\mathit{n}] = \mu_{ii} | \mathbf{h}_{\mathit{ii}}^{\mathit{H}} [\mathit{n}] \mathbf{b}_{\mathit{i}} [\mathit{n}] |^2
\end{equation}

Thus, the resulting sum rate will be:

\begin{equation}
\label{perfect}
R_{perfect} [\mathit{n}] = \sum_{j = 1}^{K} \log_2 (1 + \mu_{ii} | \mathbf{h}_{\mathit{ii}}^{\mathit{H}} [\mathit{n}] \mathbf{b}_{\mathit{i}} [\mathit{n}] |^2)
\end{equation}

\subsection{Delayed CSI Modelling}

Here we will quantize the channel vector. We will use Random Vector Quantization (RVQ) method \cite{rvq} as it is easier to analyze the efficiency of its performance. Let $\mathbf{\hat{h}}_{\mathit{ij}}$ be the quantized channel vector and $B_{ij}$ be the number of bits assigned to the channel. 

Each channel will generate a different set of codebook \cite{codebook}. Each codebook vector $c_{ij}$ is a unit norm vector which is has a zero mean and unit variance.

Let $C_{ij}$ denote the codebook matrix for a particular channel where $C_{ij} = \left[ c_{ij}^1, c_{ij}^2, \dots, c_{ij}^{2^{B_{ij}}} \right]$.

Thus the quantized channel vector will be

\begin{equation}
\mathbf{\hat{h}}_{\mathit{ij}} [\mathit{n}] = \text{ argmax}_{c_{ij}} |\mathbf{\bar{h}}_{ij}^H c_{ij}^k| \quad \forall k \in \lbrace 1, 2, \dots, 2^{B_{ij}} \rbrace
\end{equation}

For the beamforming vectors it is given by:

\begin{equation}
\mathbf{\hat{H}}_{\mathit{i}} [\mathit{n}] \mathbf{\hat{b}}_{\mathit{i}} [\mathit{n}] = 0 \text{ for } \mathit{i} \in \{1, 2, \dots, \mathit{K}\}
\end{equation}

where,

\begin{equation}
\mathbf{\hat{H}}_{\mathit{i}} [\mathit{n}] =  \left[\mathbf{\hat{h}}_{1 \mathit{i}} [\mathit{n}], \dots, \mathbf{\hat{h}}_{\mathit{(i - 1)i}} [\mathit{n}], \mathbf{\hat{h}}_{\mathit{(i + 1)i}} [\mathit{n}], \dots, \mathbf{\hat{h}}_{\mathit{Ki}} [\mathit{n}] \right]^{\mathit{H}}
\end{equation}

The nullspace is calculated in a similar way as explained before. The SINR would be:

\begin{equation}
SINR_{\mathit{i}} [\mathit{n}] = \Bigg( 1 + \frac{ \mu_{ii} | \mathbf{h}_{\mathit{ii}}^{\mathit{H}} [\mathit{n}] \mathbf{\hat{b}}_{\mathit{i}} [\mathit{n}] |^2 }{ 1 + \sum_{j = 1, j \neq i}^{K} \mu_{ij}  |\mathbf{h}_{\mathit{ij}}^{\mathit{H}} [\mathit{n}] \mathbf{\hat{b}}_{\mathit{j}} [\mathit{n}] |^2} \Bigg)
\end{equation}

Thus, the resulting sum rate will be:

\begin{equation}
\label{delayed}
R_{delayed} [\mathit{n}] = \sum_{j = 1}^{K} \log_2 \Bigg( 1 + \frac{ \mu_{ii} | \mathbf{h}_{\mathit{ii}}^{\mathit{H}} [\mathit{n}] \mathbf{\hat{b}}_{\mathit{i}} [\mathit{n}] |^2 }{ 1 + \sum_{j = 1, j \neq i}^{K} \mu_{ij}  |\mathbf{h}_{\mathit{ij}}^{\mathit{H}} [\mathit{n}] \mathbf{\hat{b}}_{\mathit{j}} [\mathit{n}] |^2} \Bigg)
\end{equation}

\section{Upper Bound Proof of Minimal Feedback Period}

The expected rate loss \cite{MFP} is given by:

\begin{align}
\label{loss}
\mathbb{E} \left( \Delta R_{i} [\mathit{n}] \right) = \mathbb{E} \left( R_{perfect} [\mathit{n}] \right) - \mathbb{E} \left( R_{delayed} [\mathit{n}] \right) 
\end{align}

Substituting the values of achievable rate in perfect CSI model (Equation \ref{perfect}) and  delayed CSI model (Equation \ref{delayed}) for channel $i$ in Equation \ref{loss}, we get:

\begin{align*}
\left( \Delta R_{i} [\mathit{n}] \right) &= \mathbb{E} \log \bigg( 1 + \mu_{ii} | \mathbf{h}_{\mathit{ii}}^{\mathit{H}} [\mathit{n}] \mathbf{b}_{\mathit{i}} [\mathit{n}] |^2 \bigg) \\
&- \mathbb{E} \log \Bigg( 1 + \frac{ \mu_{ii} | \mathbf{h}_{\mathit{ii}}^{\mathit{H}} [\mathit{n}] \mathbf{\hat{b}}_{\mathit{i}} [\mathit{n}] |^2 }{ 1 + \sum_{j = 1, j \neq i}^{K} \mu_{ij}  |\mathbf{h}_{\mathit{ij}}^{\mathit{H}} [\mathit{n}] \mathbf{\hat{b}}_{\mathit{j}} [\mathit{n}] |^2} \Bigg)
\end{align*}

Opening up the log brackets, we get:

\begin{align*}
&= \mathbb{E} \log \bigg( 1 + \mu_{ii} | \mathbf{h}_{\mathit{ii}}^{\mathit{H}} [\mathit{n}] \mathbf{b}_{\mathit{i}} [\mathit{n}] |^2 \bigg) \\
&- \mathbb{E} \log \bigg( 1 + \mu_{ii} | \mathbf{h}_{\mathit{ii}}^{\mathit{H}} [\mathit{n}] \mathbf{\hat{b}}_{\mathit{i}} [\mathit{n}] |^2 + \sum_{i = 1, i \neq j}^{K} \mu_{ij}  |\mathbf{h}_{\mathit{ij}}^{\mathit{H}} [\mathit{n}] \mathbf{\hat{b}}_{\mathit{j}} [\mathit{n}] |^2 \bigg) \\
&+ \mathbb{E} \log \Bigg( 1 + \sum_{i = 1, i \neq j}^{K} \mu_{ij}  |\mathbf{h}_{\mathit{ij}}^{\mathit{H}} [\mathit{n}] \mathbf{\hat{b}}_{\mathit{j}} [\mathit{n}] |^2 \Bigg)
\end{align*}

Since, $| \mathbf{h}_{\mathit{ii}}^{\mathit{H}} [\mathit{n}] \mathbf{\hat{b}}_{\mathit{i}} [\mathit{n}] |^2 \geq 0$, the upper bound of $\log \bigg( 1 + \mu_{ii} | \mathbf{h}_{\mathit{ii}}^{\mathit{H}} [\mathit{n}] \mathbf{\hat{b}}_{\mathit{i}} [\mathit{n}] |^2 \bigg)$ can be written as $\log \bigg( 1 + \mu_{ii} | \mathbf{h}_{\mathit{ii}}^{\mathit{H}} [\mathit{n}] \mathbf{\hat{b}}_{\mathit{i}} [\mathit{n}] |^2 \bigg)$. This will lead to a simplified bound :

\begin{align*}
&\leq \mathbb{E} \log \bigg( 1 + \mu_{ii} | \mathbf{h}_{\mathit{ii}}^{\mathit{H}} [\mathit{n}] \mathbf{b}_{\mathit{i}} [\mathit{n}] |^2 \bigg) \\
&- \mathbb{E} \log \bigg( 1 + \mu_{ii} | \mathbf{h}_{\mathit{ii}}^{\mathit{H}} [\mathit{n}] \mathbf{\hat{b}}_{\mathit{i}} [\mathit{n}] |^2 \bigg) \\
&+ \mathbb{E} \log \Bigg( 1 + \sum_{i = 1, i \neq j}^{K} \mu_{ij}  |\mathbf{h}_{\mathit{ij}}^{\mathit{H}} [\mathit{n}] \mathbf{\hat{b}}_{\mathit{j}} [\mathit{n}] |^2 \Bigg)
\end{align*}

The beamforming vectors $\mathbf{b}_{\mathit{i}} [\mathit{n}] $ and $\mathbf{\hat{b}}_{\mathit{i}} [\mathit{n}]$ are unit vectors which are isotropically distributed, and they are independent from the channel vector $\mathbf{h}_{\mathit{ij}} [\mathit{n}]$. Thus the terms $ \mathbb{E} \log \bigg( 1 + \mu_{ii} | \mathbf{h}_{\mathit{ii}}^{\mathit{H}} [\mathit{n}] \mathbf{b}_{\mathit{i}} [\mathit{n}] |^2 \bigg)$ and $\mathbb{E} \log \bigg( 1 + \mu_{ii} | \mathbf{h}_{\mathit{ii}}^{\mathit{H}} [\mathit{n}] \mathbf{\hat{b}}_{\mathit{i}} [\mathit{n}] |^2 \bigg)$ will cancel each other out.

\begin{align*}
\mathbb{E} \left( \Delta R_{i} [\mathit{n}] \right) \leq \mathbb{E} \log \Bigg( 1 + \sum_{i = 1, i \neq j}^{K} \mu_{ij}  |\mathbf{h}_{\mathit{ij}}^{\mathit{H}} [\mathit{n}] \mathbf{\hat{b}}_{\mathit{j}} [\mathit{n}] |^2 \Bigg)
\end{align*}

Using Jensen's inequality, the expectation term can be taken up inside leading to this:

\begin{align}
\label{this}
\mathbb{E} \left( \Delta R_{i} [\mathit{n}] \right) \leq \log \Bigg( 1 + \sum_{i = 1, i \neq j}^{K} \mu_{ij} \, \mathbb{E} \left( |\mathbf{h}_{\mathit{ij}}^{\mathit{H}} [\mathit{n}] \mathbf{\hat{b}}_{\mathit{j}} [\mathit{n}] |^2 \right) \Bigg)
\end{align}

The upper bound on $\mathbb{E} \left( |\mathbf{h}_{\mathit{ij}}^{\mathit{H}} [\mathit{n}] \mathbf{\hat{b}}_{\mathit{j}} [\mathit{n}] |^2 \right)$ is given by:

\begin{align*}
&\mathbb{E} \left( |\mathbf{h}_{\mathit{ij}}^{\mathit{H}} [\mathit{n}] \mathbf{\hat{b}}_{\mathit{j}} [\mathit{n}] |^2 \right) \\
&\leq 1 - \epsilon_{\mathit{ij}}^2 + \epsilon_{\mathit{ij}}^2 2^{B_{ij}} \beta \bigg( 2^{B_{ij}}, \frac{M}{M - 1} \bigg) \frac{M}{M - 1}
\end{align*}

Jindal has given a detailed proof for the above beta-distribution based bound \cite{Jindal}. It is known that, $ \beta(x, y) = \frac{\Gamma (x) \Gamma (y)}{\Gamma (x + y)}$. If $x \gg y$, it can be approximated as : $ \beta(x, y) \approx \Gamma (y) x^{-y}$. Thus, 

\begin{align*}
&\mathbb{E} \left( |\mathbf{h}_{\mathit{ij}}^{\mathit{H}} [\mathit{n}] \mathbf{\hat{b}}_{\mathit{j}} [\mathit{n}] |^2 \right) \\
&\leq 1 - \epsilon_{\mathit{ij}}^2 + \epsilon_{\mathit{ij}}^2 2^{B_{ij}} \Gamma \left( \frac{M}{M - 1} \right) 2^{\frac{-B_{ij}M}{M - 1}} \frac{M}{M - 1}
\end{align*}

Substituting $\mathbb{E} \left( |\mathbf{h}_{\mathit{ij}}^{\mathit{H}} [\mathit{n}] \mathbf{\hat{b}}_{\mathit{j}} [\mathit{n}] |^2 \right)$ in equation \ref{this}, we get:

\begin{align*}
\Delta_R \approx \log \Bigg( 1 + \sum_{i = 1, i \neq j}^{K} \mu_{ij} \, &\bigg( 1 - \epsilon_{\mathit{ij}}^2 + \epsilon_{\mathit{ij}}^2 2^{B_{ij}} \\
&\Gamma \left( \frac{M}{M - 1} \right) 2^{\frac{-B_{ij}M}{M - 1}} \frac{M}{M - 1} \bigg) \Bigg)
\end{align*}

For $M > 1$, we can ignore the higher order terms of $2^{\frac{-B_{ij}M}{M - 1}}$ in the summation. Using the identity $N \Gamma (N) = \Gamma (N + 1)$ this can be further approximated as

\begin{align*}
\Delta_R &\approx \log \Bigg( 1 + \bigg( \sum_{i = 1, i \neq j}^{K} \mu_{ij} (1 - \epsilon_{\mathit{ij}}^2) \bigg) \\
&+ \Gamma \left( \frac{2M - 1}{M - 1} \right)M \bigg( \sum_{i = 1, i \neq j}^{K} \mu_{ij} \epsilon_{\mathit{ij}}^2 2^{\frac{-B_{ij}}{M - 1}} \bigg) \Bigg)
\end{align*}

The log is monotonic function and $B_{ij}$ are the only optimization parameters, the above problem can be reduced into:

\begin{align}
\label{obj}
\min_{B_{ij}}  \bigg( \sum_{i = 1, i \neq j}^{K} \mu_{ij} \epsilon_{\mathit{ij}}^2 2^{\frac{-B_{ij}}{M - 1}} \bigg) 
\end{align}

where $\sum_{i = 1, i \neq j}^{K} B_{ij} = B_{s}$. The total number of bits which can be allotted between the vectors is denoted as $B_s$. Thus $B_s$ can act as an constraint on the minimization problem.

The lagrange function for the minimization of the objective function (Equation \ref{obj}) can be formed as:

\begin{align}
f(B_{ij}, \lambda) = \Bigg( \sum_{i = 1, i \neq j}^{K} \mu_{ij} \epsilon_{\mathit{ij}}^2 2^{\frac{-B_{ij}}{M - 1}} \Bigg) + \lambda \Bigg(\sum_{i = 1, i \neq j}^{K} B_{ij} - B_{s} \Bigg)
\end{align}

Putting $\frac{\partial f}{\partial B_{ij}} = 0$, we get:

\begin{align}
\label{lambda}
\lambda = \frac{ \mu_{ij} \epsilon_{\mathit{ij}}^2 2^{\frac{-B_{ij}}{M - 1}} \log 2 }{M - 1}
\end{align}

Taking out the $B_{ij}$ term from equation \ref{lambda} , we get:

\begin{align}
\label{e2}
B_{ij} = (M - 1) \log \bigg( \frac{\mu_{ij} \epsilon_{\mathit{ij}}^2 \log 2}{(M - 1) \lambda} \bigg)
\end{align}

Putting $\frac{\partial f}{\partial \lambda} = 0$, we get:

\begin{align}
\label{derive}
\sum_{i = 1, i \neq j}^{K} B_{ij} = B_{s}
\end{align}

Substituting $B_{ij}$, in the equation \ref{derive}, we get:
\begin{align}
\label{e1}
\sum_{i = 1, i \neq j}^{K} \Bigg[ (M - 1) \log \bigg( \frac{\mu_{ij} \epsilon_{\mathit{ij}}^2 \log 2}{(M - 1) \lambda} \bigg) \Bigg] = B_{s}
\end{align}

Taking out the $\lambda$ term from equation \ref{e1} gives:

\begin{align}
\lambda = \log (2) \, 2 ^{\frac{-B_{s}}{(M - 1)^2}} \left( \prod_{i = 1, i \neq j}^{K} \mu_{ij} \epsilon_{\mathit{ij}}^2 \right)^{\frac{1}{M - 1}}
\end{align}

Substituting the value of $\lambda$ in the equation \ref{e2} gives:

\begin{align}
\label{omg}
B_{ij} = (M - 1) \log \Bigg( \frac{\mu_{ij} \epsilon_{\mathit{ij}}^2 \log (2)}{\log (2) \, 2 ^{\frac{-B_{s}}{(M - 1)^2}} \left( \prod_{i = 1, i \neq j}^{K} \mu_{ij} \epsilon_{\mathit{ij}}^2 \right)^{\frac{1}{M - 1}}} \Bigg)
\end{align}

Simplifying equation \ref{omg} gives:

\begin{align}
\label{haha}
B_{ij} = \frac{B_{s}}{M - 1} + (M - 1) \Bigg[ \log \left( \frac{\mu_{ij} \epsilon_{\mathit{ij}}^2}{\prod_{i = 1, i \neq j}^{K} \left( \mu_{ij} \epsilon_{\mathit{ij}}^2 \right)^{\frac{1}{M - 1}}}  \right) \Bigg]
\end{align}

Thus we get the optimal value of $B_{ij}$ in the MFP scheme which effectively minimizes the objective function given in the equation \ref{obj}.

\section{Calculations of Adaptive Feedback Period}

The expected rate loss is defined as:

\begin{align*}
\left( \Delta R_{i} [\mathit{n}] \right) &= \mathbb{E} \log \bigg( 1 + \mu_{ii} | \mathbf{h}_{\mathit{ii}}^{\mathit{H}} [\mathit{n}] \mathbf{b}_{\mathit{i}} [\mathit{n}] |^2 \bigg) \\
&- \mathbb{E} \log \Bigg( 1 + \frac{ \mu_{ii} | \mathbf{h}_{\mathit{ii}}^{\mathit{H}} [\mathit{n}] \mathbf{\hat{b}}_{\mathit{i}} [\mathit{n}] |^2 }{ 1 + \sum_{j = 1, j \neq i}^{K} \mu_{ij}  |\mathbf{h}_{\mathit{ij}}^{\mathit{H}} [\mathit{n}] \mathbf{\hat{b}}_{\mathit{j}} [\mathit{n}] |^2} \Bigg)
\end{align*}

The upper bound of the rate loss \cite{awful} is defined as :

\begin{equation}
\overbar{\Delta R_{i}} = \log (1 + g(B_{ij}, \omega_{ij}))
\end{equation}

where :

\begin{equation}
g(B_{ij}, \omega_{ij}) = \displaystyle\sum_{j = 1, j \neq i}^K  \mu_{ij} \left\{\epsilon_{ij}^{2(\omega_{ij} - 1)} \left(\dfrac{M}{M - 1} 2^{-\frac{\omega_{ij}B_{ij}^t}{(M - 1)T}} - 1\right) + 1\right\}
\end{equation}

with the constraint:

\begin{equation}
\sum_{\substack{
   j = 1 \\
   j \neq i
  }}^{K}
 B_{ij}^{t} = B_s
\end{equation}

Here, $\omega_{ij}$ is the time period whenever the quantization vector would be updated and $B_{ij}^{t}$ are the total bits assigned to a particular channel.

If a time frame is a multiple of $\omega_{ij}$, or $n = k \omega_{ij}$, then at that $n^{th}$ instant we will feed back the quantized channel vector. If $n \neq k \omega_{ij}$, the channel will use the recently quantized vector and thus it will not form corresponding beamforming vector at that particular instant. In other words, MFP has the $\omega_{ij} = 1$, but in AFP both $\omega_{ij}$ and $B_{ij}^{t}$ needs to be calculated. In this section, we will show how to construct a Jacobian and Hessian matrix which can be used by any optimization programming library to determine these parameters.

The constrained optimization can be changed into unconstrained optimization by substitution. The constrained can be altered as:

\begin{equation}
B_{Kj}^{t} = B_T - \sum_{\substack{
   j = 1 \\
   j \neq i
  }}^{K - 1}
 B_{ij}^{t}
\end{equation}

Substituting $B_{Kj}^{t}$ in $g(B_{ij}, \omega_{ij})$ we get:

\begin{align*}
g = \left(\displaystyle\sum_{j = 1, j \neq i}^{K - 1} \mu_{ij} \left\{\epsilon_{ij}^{2(\omega_{ij} - 1)} \left(\dfrac{M}{M - 1} 2^{-\frac{\omega_{ij}B_{ij}^t}{(M - 1)T}} - 1\right) + 1\right\} \right) \\
+ \mu_{Kj} \left\{\epsilon_{Kj}^{2(\omega_{Kj} - 1)} \left(\dfrac{M}{M - 1} 2^{-\frac{\omega_{Kj} \left(B_T - \sum_{j = 1, j \neq i}^{K - 1}B_{ij}^{t}\right)}{(M - 1)T}} - 1\right) + 1\right\}
\end{align*}

Jacobian matrix $\mathbf{J}$ of the optimization function will be:

\begin{equation*}
\mathbf{diag}(\mathbf{J}) = \left[ \frac{\partial g}{\partial\omega_{i1}}, \frac{\partial g}{\partial\omega_{i2}}, \cdots, \frac{\partial g}{\partial\omega_{iK}}, \frac{\partial g}{\partial B_{i1}}, \cdots, \frac{\partial g}{\partial B_{i(K - 1)}} \right]
\end{equation*}

where,
\begin{align*}
\frac{\partial g}{\partial\omega_{ij}} = \mu_{ij}\Bigg[2 \epsilon_{ij}^{2(\omega_{ij} - 1)} \log(\epsilon_{ij})\left(\frac{M}{M-1} 2^{-\frac{\omega_{ij}B_{ij}^t}{(M - 1)T}} - 1\right) \\
 - B_{ij}^{t} \log(2) \, \epsilon_{ij}^{2(\omega_{ij} - 1)} \frac{M}{(M - 1)^2 T} 2^{-\frac{\omega_{ij}B_{ij}^t}{(M - 1)T}} \Bigg]
\end{align*}
Similarly, other entries of the Jacobian matrix can be calculated.

Hessian matrix $\mathbf{H}$ of the given optimization function is:
\begin{align*}
\mathbf{H} = 
\begin{bmatrix}
\frac{\partial^2 g}{\partial \omega_{i1}^2} & \cdots & \frac{\partial^2 g}{\partial B_{i(K-1)} \partial \omega_{i1} } \\
\frac{\partial^2 g}{\partial \omega_{i1} \partial \omega_{i2}} & \cdots & \frac{\partial^2 g}{\partial B_{i(K-1)} \partial \omega_{i2} } \\
\vdots & \ddots & \vdots \\
\frac{\partial^2 g}{\partial \omega_{i1} \partial \omega_{iK}} & \cdots & \frac{\partial^2 g}{\partial B_{i(K-1)} \partial \omega_{iK} } \\
\frac{\partial^2 g}{\partial \omega_{i1} \partial B_{i1}} & \cdots & \frac{\partial^2 g}{\partial B_{i(K-1)} \partial B_{i1} } \\
\frac{\partial^2 g}{\partial \omega_{i1} \partial B_{i2}} & \cdots & \frac{\partial^2 g}{\partial B_{i(K-1)} \partial B_{i2} } \\
\vdots & \ddots & \vdots \\
\frac{\partial^2 g}{\partial \omega_{i1} \partial B_{i(i - 1)}} & \cdots & \frac{\partial^2 g}{\partial B_{i(K-1)}^2 } \\
\end{bmatrix}
\end{align*}
Clearly, this matrix is symmetric and has a dimensions $(M + K - 1) \times (M + K - 1)$. The entries of the hessian matrix are:

\begin{align*}
\frac{\partial^2 g}{\partial b_{ij}\omega_{ij}} =&  - \frac{\log(2) M}{(M - 1)^3 T^2} \times \\
&\sum_{\substack{l = 1, l \neq i}}^{K - 1} \mu_{ij} \epsilon_{ij}^{2(\omega_{ij} - 1)} 2^{-\frac{\omega_{ij}B_{ij}^t}{(M - 1)T}} \\
&\bigg[ 2 (M - 1) \omega_{ij} T \log(\epsilon_{ij}) \\
&- \omega_{ij}B_{ij}^t \text{ log }(2) + (M - 1)T \bigg]
\end{align*}
Similarly, other entries of the hessian matrix can be calculated.

Through simulation , we found out that the Hessian matrix obtained is symmetric and positive definite. Thus the $g(B_{ij}, \omega_{ij})$  is convex with a unique minima.

We set the values, $v_{12} = 9 \, kmph$, $v_{13} = 8 \, kmph$, $M = 3$, $T_s = 5 \, ms$, $T = 30$. The convergence of $g(B_{ij}, \omega_{ij})$ is plotted against number of iterations. We used here Newton Conjugate Gradient method for the convex optimization. The convergence is obtained within 10 iterations.

\section{Simulation Results and Discussions}

\begin{table}[h]
\centering
\caption{Simulation Parameters}
\label{simulation}
\begin{tabular}{|c|c|}
\hline
\multicolumn{1}{|c|}{\textbf{Parameter}} & \multicolumn{1}{c|}{\textbf{Value}} \\ \hline
$M$ & 3 \\
$T_s$ & 5 ms \\
$T$ & 30 \\
$v_1$ & 10 kmph \\
$v_2$ & 9 kmph \\
$v_3$ & 8 kmph \\
$B_s$ & 20 bits\\
$\mu_{11}$ & 10 dB to 19 dB \\
$\mu_{12}$ & $ \mu_{11} - 2$ \\
$\mu_{13}$ & $ \mu_{11} - 3$ \\ 
\hline
\end{tabular}
\end{table}

The simulation parameters are summarized in Table \ref{simulation}. The simulation of allocation schemes is plotted with power constraints against mean spectral efficiency (mean sum-rate metric). The power constraints are looped throughout in order to plot the simulation result graph of average mean sum rate vs power constraint of the first receiver ($\mu_{11}$). Here $\mu_{11}$ is looped over from 10 dB to 19 dB with $\mu_{12} = \mu_{11} - 2$ and $\mu_{13} = \mu_{11} - 3$.

The simulation is repeated over 500 times and different set of codebooks are used every $50^{th}$ time. The simulation is performed for three bit allocation strategies:

\begin{enumerate}
\item Adaptive Feedback Period (AFP) bit and feedback rate allocation
\item Minimal Feedback Period (MFP) adaptive bit allocation
\item Minimal Feedback Period (MFP) equal bit allocation
\end{enumerate}

The simulation of the model was done with the help of SciPy libraries. The channel vectors were constructed using random complex vectors, which followed a Gaussian distribution of $\mathcal{N}(0, \mathbf{I})$. Similarly, the desired beamforming vectors were formed for the channels. The number of bits required for quantization of beamforming vectors was calculated from equation \ref{haha} for the MFP scheme. The number of bits and feedback period for the AFP scheme was determined through the joint optimization of the equation 27. The SINR for the schemes were calculated which were averaged over the entire time period. The mean of the resulting values of the channels was taken and plotted in the graph, where the power constraint of the first channel was varied. 

\section{Conclusion}
\begin{figure}[h]
\centering
\includegraphics[width=3.2 in]{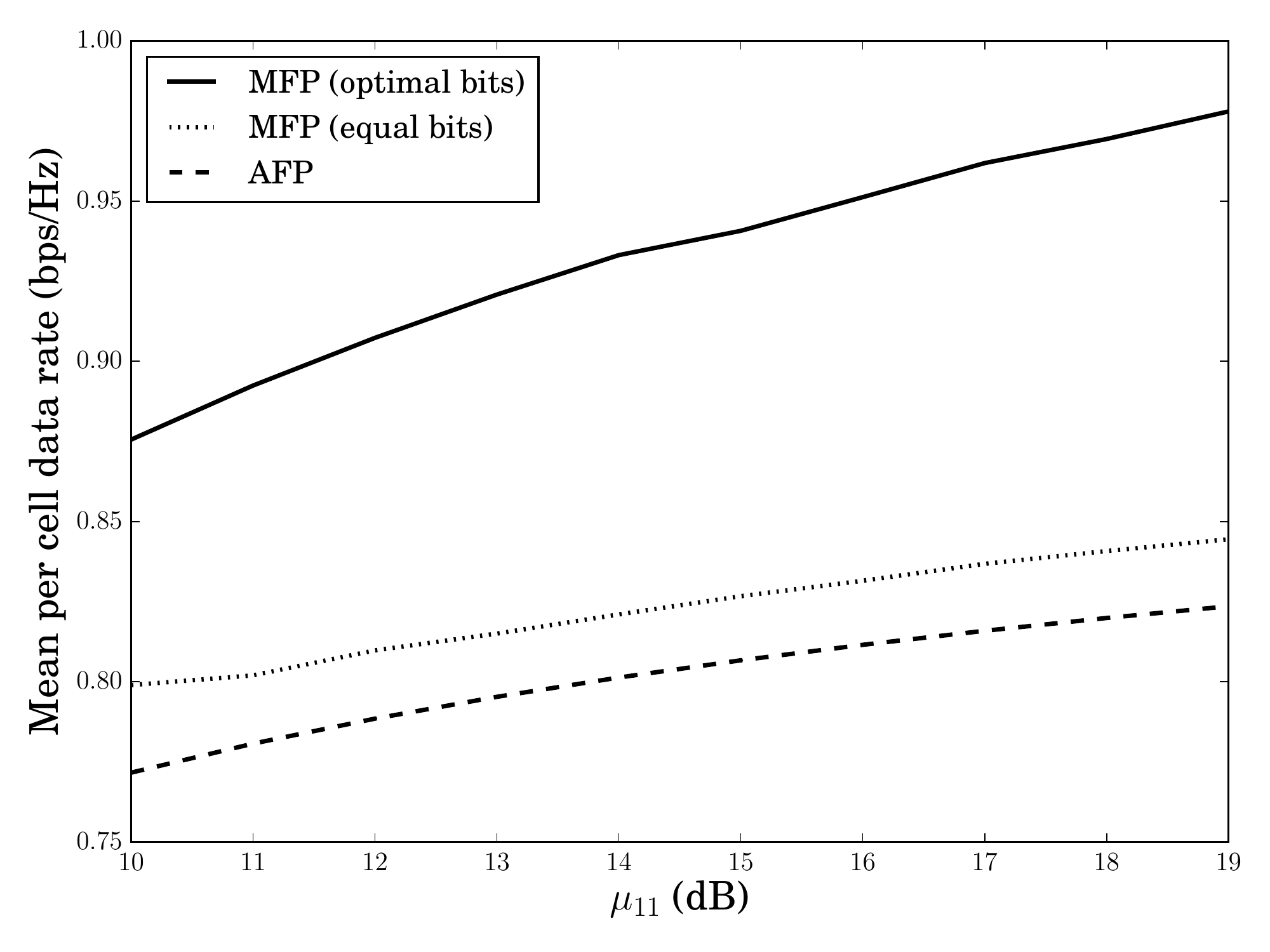}
\caption{Simulation results of the bit allocation schemes. Spectral efficiency of the model is plotted against the power constraint of the channel by using different bit allocation schemes.}
\label{final}
\end{figure}
This paper outlines the MFP scheme based on the bit allocation parameter. 
The simulation results are plotted in the Figure \ref{final}. Here, adaptive MFP outperforms the AFP scheme. The performance gap is more indicative as power constraint increases. We see that simpler MFP scheme and solution surpasses the more complex AFP scheme, in which solutions are obtained through numerical optimization methods. 

MFP scheme could be applied with the systems similar to the Gauss Markov fading models to allocate the bits in mobile networking.

\bibliographystyle{IEEEtran}
\bibliography{IEEEexample}

\end{document}